\documentclass[aps,prb,amsmath,amssymb,twocolumn,floatfix,10pt,showpacs,superscriptaddress]{revtex4-1}

\usepackage{graphicx}
\usepackage{color}
\usepackage{epstopdf}

\begin{document}

\title{Entanglement entropy scaling in the bilayer Heisenberg spin system }
\author{Johannes Helmes}
\affiliation{Institut f\"ur Theoretische Festk\"orperphysik, JARA-FIT and JARA-HPC, RWTH Aachen University, 52056 Aachen, Germany}
\affiliation{Institut f\"ur Theoretische Physik, Universit\"at zu K\"oln, 50923 K\"oln, Germany}
\author{Stefan Wessel}
\affiliation{Institut f\"ur Theoretische Festk\"orperphysik, JARA-FIT and JARA-HPC, RWTH Aachen University, 52056 Aachen, Germany}
\date{\today}

\begin{abstract}
We examine the entanglement properties of the spin-half Heisenberg model on the two-dimensional square-lattice bilayer based on 
quantum Monte Carlo calculations of the second R\'enyi entanglement entropy.
In particular, we extract the dominant area-law contribution to the bipartite entanglement entropy that shows a non-monotonous behavior upon increasing the inter-layer exchange interaction: a local maximum in the area-law coefficient is located at the quantum critical point separating the antiferromagnetically ordered region from the disordered dimer-singlet regime. Furthermore, we consider   subleading logarithmic corrections to the R\'enyi entanglement entropy scaling.
Employing different subregion shapes, we isolate the logarithmic corner term  from the logarithmic contribution due to Goldstone modes that is  found to be enhanced  in the  limit of decoupled layers.  At the quantum critical point , we estimate a contribution of $0.016(1)$ due to each $90^{\circ}$ corner.  This corner term at the SU(2) quantum critical point  deviates from the Gaussian theory value, while it compares well with recent numerical linked cluster calculations on the bilayer model. 
\end{abstract}

\maketitle

\section{Introduction}
In recent years, the study of the entanglement between subregions has been identified as a promising theoretical probe of the quantum correlations in strongly correlated many-body quantum systems. Of particular interest in this respect is the identification of universal contributions to the scaling of the entanglement entropy with the size of the boundary that separates the subregions. The dominant contribution to the bipartite entanglement entropy in quantum systems beyond one spatial dimension generically scales proportional to the area of the boundary that separates two subregion~\cite{bombelli86, srednicki93, eisert08} (however, there are several notable exceptions~\cite{wolf06, gioev06, fursaev06, lai13}). Accounting for all microscopic degrees of freedom along the boundary, the corresponding area-law scaling coefficient is non-universal, so that the extraction of subleading corrections beyond the  area law is required in order to explore possible universal contributions. 
For one-dimensional quantum systems, a detailed understanding of the entanglement entropy scaling
has  been well established. In particular, for gapped one-dimensional systems a finite entanglement entropy relates to the one-dimensional version of the area law, while quantum critical systems  
exhibit a logarithmically divergent (in the subsystem's length) entanglement entropy with the scaling coefficient determined by the central charge of the corresponding conformal field theory~\cite{vidal03, calabrese04, calabrese09}. 

Advancing beyond the one-dimensional case, several 
subleading contributions to the entanglement entropy of two-dimensional quantum systems have been considered recently (cf. Ref.~\onlinecite{inglis13} for a recent overview):  E.g., in gapped systems, a universal finite subleading contribution relates to the topology of the entangled state in topologically ordered phases, and gives rise to a topological entanglement entropy constant  e.g. for spin liquid ground states~\cite{kitaev06,levin06}.
Subleading logarithmic (in the boundary length) terms have been identified in continuous-symmetry broken phases~\cite{kallin11} and were suggested to derive in a universal way from the presence of Goldstone modes through the restoration of the symmetry on finite geometries~\cite{metlitski11}.  
In addition, subleading  logarithmic  corrections appear due to specific 
geometric features such as corners or vertices on the subregion boundary, and were found to depend on universal characteristics e.g. at several standard examples of conformal two-dimensional quantum critical points ~\cite{ardonne04,fradkin06, casini07, kallin11, singh12, kallin13, inglis13, myers11, kallin14}.  
Much of the recent progress in identifying universal contributions to the entanglement entropy has been made possible by combining analytical and field-theoretical results with advanced numerical methods to extract entanglement properties of specific interacting quantum lattice models employing e.g. series expansion methods~\cite{singh12},  density matrix renormalization group~\cite{white92} calculations~\cite{yan10, depenbrock12, jiang12}, as well as quantum Monte Carlo (QMC) simulations~\cite{hastings10, kallin11, humeniuk12}.
 
Of particular interest in this respect are model systems  exhibiting a quantum phase transition that separates distinct quantum phases. Several recent studies considered the two-dimensional transverse-field Ising model, which allows  to probe the entanglement properties in a system with well-characterized phases and separated by a quantum critical point that belongs to  the three-dimensional Ising model universality class~\cite{humeniuk12, kallin13,inglis13}. Another basic quantum spin model that exhibits a well-characterized quantum phase transition is the spin-half Heisenberg model on the square-lattice bilayer. In contrast to the transverse-field Ising model, this model exhibits a \emph{continuous} internal symmetry, and correspondingly, it  has a quantum critical point of (three-dimensional) O(3) universality~\cite{wang06}. 
Here, we apply QMC simulations to compute the R\'enyi~\cite{renyi61} entropy-based bipartite entanglement measure in the spin-half Heisenberg model on the  square-lattice bilayer and study its entanglement properties both in the magnetically ordered and disordered phase as well as at the quantum critical point.  In addition to analyzing the dominant area-law contribution to the entanglement entropy in this model upon varying the interaction ratio between the inter- and intralayer exchange interactions, we also analyze the contributions to the entanglement entropy that arise due to  Goldstone modes as well as corner contributions and which are considered to be universal.
Recently,  the Heisenberg bilayer model was examined at its quantum critical point using a numerical linked cluster expansion approach with a focus towards the contribution of $90^{\circ}$ corners in the subregion boundary to the  entanglement entropy~\cite {kallin14}, and we will relate to these results further below. 
The remainder of this paper is organized as follows: in the next section we introduce the model and subregions under consideration as well as the numerical method that we employ to calculate the entanglement entropy. Then, in Sec. III, we present our numerical data for the entanglement properties and possible universal contributions.  Sec. IV  provides a final discussion of our findings. 

\section{Model and Method}
In the following, we consider the spin-half Heisenberg model on the square-lattice bilayer, described by the Hamiltonian
\begin{equation}
 H=J\sum_{\langle i,j\rangle} \left(\mathbf{S}_ {i,1}\cdot\mathbf{S}_ {j,1} +  \mathbf{S}_ {i,2}\cdot\mathbf{S}_ {j,2}\right)+J_{\perp}\sum_i \mathbf{S}_ {i,1}\cdot\mathbf{S}_ {i,2}
\end{equation}
where $i$ denotes the $i$-th unit cell containing two spin-half degrees of freedom and $J$ ($J_\perp$) the intralayer (interlayer) exchange interaction. 
We denote by $g=J_\perp/J$ the ratio of the interlayer to the intralayer exchange interactions.
From previous studies, it is well established, that the critical coupling ratio $g_c=2.5220(1)$ separates the low-$g$ antiferromagnetically ordered phase from the high-$g$ magnetically disordered dimer spin singlet phase. The quantum phase transition at $g=g_c$ belongs to the three-dimensional O(3) universality class, and has been well characterized~\cite{wang06}.

The bilayer system will be considered in the following as a square-lattice model with a two-spin unit cell, one corresponding to each layer.  
For the quantum Monte Carlo simulations, we then consider finite lattices of linear extend $L$, containing $L^2$ square lattice sites (i.e. bilayer unit cells) and $2L^2$ spins, with periodic boundary conditions employed in both lattice directions, corresponding to a toroidal simulation cell.  

In order to study the entanglement properties of this model within both phases and at criticality, we employ QMC simulations to extract the R\'enyi entropy-based bipartite entanglement measure
\begin{equation}
 S_\alpha(A)=\frac{1}{1-\alpha}\ln\mathrm{Tr}[(\rho_A)^\alpha]
\end{equation}
between a subregion $A$ and its complement in terms of the reduced density matrix $\rho_A$ for subregion A. Recently, various QMC methods have been devised to calculate $S_\alpha$.  In particular, we focus here on the second R\'enyi entropy $S_2$, which we calculate using the extended ensemble sampling approach based on the replica trick~\cite{hastings10, melko10} within the stochastic series expansion~\cite{sandvik99} QMC representation of Ref.~\onlinecite{humeniuk12}. This approach allows us to efficiently obtain $S_2$ from a direct calculation at a chosen inverse temperature $\beta$, without the need to perform any thermodynamic integration or extended thermal ensemble  sampling~\cite{inglis13a}. Furthermore, in order to probe for ground state properties, we linearly scale $\beta J=L$ with the system size $L$, cf.  Ref.~\onlinecite{humeniuk12}. Details on the QMC method, in particular concerning the use of the ``increment trick''~\cite{hastings10} to successively calculate $S_2(A)$ upon growing the 
subregion $A$, may be found in Ref.~\onlinecite{humeniuk12}. 
\begin{figure}[!t]
\includegraphics[width=0.9\linewidth]{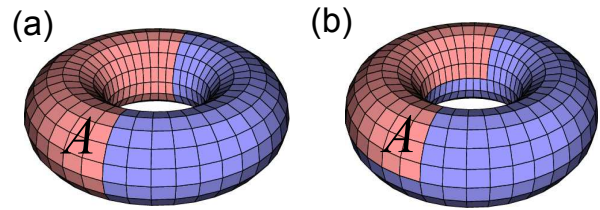}
\caption{(Color online) Different subregions $A$ considered from the $L\times L$ toroidal QMC simulation lattice: (a) $L/2 \times L$ strip subregion with smooth boundaries, (b) $L/2 \times L/2$ square subregion with four $90^{\circ}$ corners. In the figure each lattice site represents a bilayer unit cell containing two spin-half degrees of freedome.}
\label{fig:sr}
\end{figure}

In our simulations, we considered two specifically shaped subregion types (cf. Fig.~\ref{fig:sr}) : (a) a bipartition of the  toroidal simulation lattice into two equally sized cylindrical strips of size $L/2 \times L$: the total circumference of the subregion boundary then equals $l=2\times L$. (b) a square subregion of size $L/2 \times L/2$: the circumference of the subregion boundary is given as $l=4\times L/2=2L$, which also equals the average of the number of square lattice sites (i.e. bilayer unit cells) on the inside and the outside of the subregion boundary. However,  while  the strip  subregions have smooth boundaries, the square subregions each feature four $90^{\circ}$ corners along the boundary. As mentioned in the introduction, a distinct logarithmic contribution to the entanglement entropy is expected to arise due to the presence of these corners. 
We focus in the following on these two differently shaped subregions, since the linear size of both scales with the size of the simulation cell while keeping a constant aspect-ratio. Any contribution of  the entanglement entropy $S_2$ that depends only on the aspect-ratio of the subregion  thus leads to an $l$-independent contribution to $S_2$ for both considered subregions. Different functional forms for the aspect ratio-dependence of such separate shape-dependent contributions have been considered for specific model systems, and we refer to Ref.~\onlinecite{inglis13} for a recent  comparison of these functional forms for a two-dimensional quantum critical system.  

\section{Results}
We performed QMC simulations at various values of $g$ across the quantum phase transition at $g=g_c$, with $g$ ranging between $0$ and $3.5$, and for system sizes $L=8,10,12,14,16,18,20$, for which we obtained data with a relative statistical uncertainty of $O(10^{-3})$. 
\begin{figure}[!t]
\includegraphics[width=\linewidth]{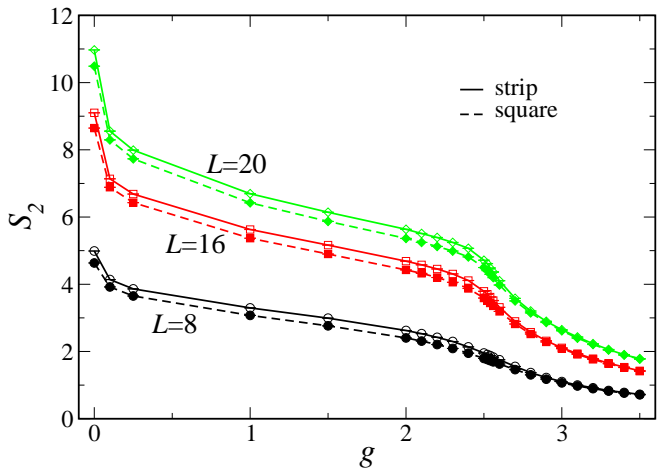}
\caption{(Color online) Second R\'enyi entropy $S_2$ as a function of the coupling ratio $g$  for the spin-half Heisenberg model on the square lattice bilayer for strip and square shaped subregions of boundary length $l=2L$ at different linear system sizes $L$.}
\label{fig:Svsj}
\end{figure}
Figure ~\ref{fig:Svsj} shows the results for $S_2$ for both the strip and square subregions for three different system sizes, $L=8,16,20$. From this figure, one notices (i) a general reduction of $S_2$ upon increasing the interlayer coupling, (ii) a shoulder developing in $S_2$ at the quantum critical point, and (iii) a smaller value of $S_2$ for the square subregion as compared to the strip subregion -- most pronounced throughout  the low-$g$, antiferromagnetically ordered region. We remark that at $g=0$  the system decouples into two independent layers, so that the values of $S_2$ for the bilayer at $g=0$ are exactly twice the values for a single square lattice layer of the same linear extend $L$. Furthermore, in the large-$g$ limit, in which the system's ground state decouples into a direct product state of singlets formed on the spin dimers in each unit cell, $S_2$ vanishes for both subregion geometries, since the subregion boundaries in both cases do not cut through these dimers. 
The reduction of $S_2$ in the large-$g$ region  makes the QMC simulations less efficient for even larger values of $g$ than those that we considered, since the ensemble exchange rate, essentially quantifying the entanglement entropy, drops significantly in this regime. 

\begin{figure}[!t]
\includegraphics[width=\linewidth]{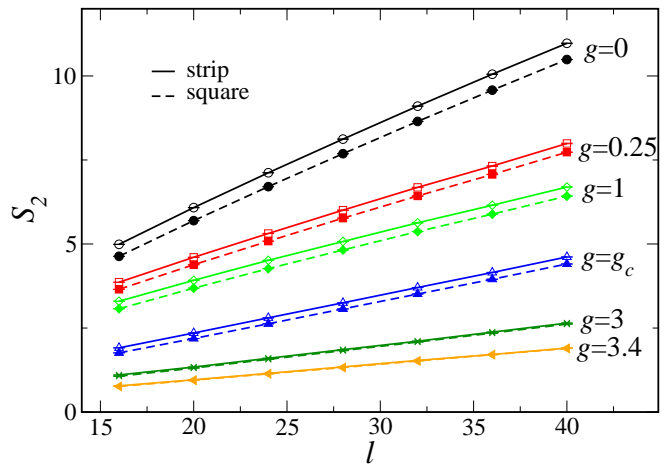}
\caption{(Color online) Second R\'enyi entropy $S_2$ as a function of the boundary length $l$ at different values of the coupling ratio $g$ for strip and square shaped subregions.}
\label{fig:Svsl}
\end{figure}

\begin{figure}[!t]
\includegraphics[width=\linewidth]{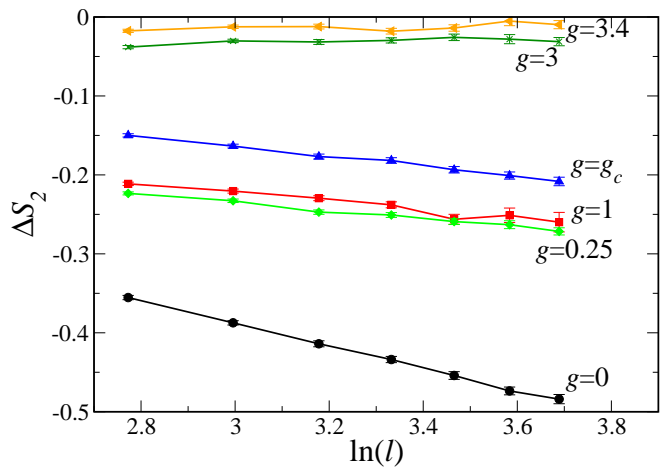}
\caption{(Color online) Difference between the second R\'enyi entropy $\Delta S_2=S_2^\mathrm{sq}-S_2^\mathrm{st}$ for square (sq) and strip (st) shaped subregions as a function of the boundary length $l$ (shown on a log-scale) at different 
values of the coupling ratio $g$.}
\label{fig:DSvsl}
\end{figure}

Fig.~\ref{fig:Svsl} shows  $S_2$ for different values of $g$  as  functions of the boundary length $l$ for both subregion geometries. The dominant linear increase of $S_2$ with $l$ is clearly visible in all cases, as is the difference in $S_2$ for the two types of subregions. This difference 
\begin{equation}\label{eq:ds2}
 \Delta S_2=S_2^\mathrm{sq}-S_2^\mathrm{st}
\end{equation}
between the entanglement entropy of the square subregion ($S_2^\mathrm{sq}$) and the strip subregion ($S_2^\mathrm{st}$) is shown for various values of $g$ in Fig.~\ref{fig:DSvsl}, and will be analyzed below to extract the corner term. 

For such a  quantitative analysis of the QMC data, we require to account for the various expected contributions to the entanglement entropy scaling with the cut length $l$.
For the entanglement entropy  a general scaling formula is given as
\begin{equation}
 S_2= a l + c\ln(l) + d + O(1/l),
\end{equation}
where  $a$ denotes the coefficient of the leading area law contribution. As mentioned in the introduction already, the prefactor $c$ of the logarithmic term may receive  contributions due to Goldstone modes e.g. in the antiferromagnetically ordered region with spontaneously broken SU(2) symmetry  as well as due to  corner in e.g. the boundaries of the square subregions. Furthermore, we include a constant term ($d$). As noted in the introduction,  a finite constant contribution to $S_2$ is  expected e.g. in topologically ordered systems with a finite topological entanglement entropy. However, the gapped phase for $g>g_c$ in the present model is topologically trivial, being adiabatically linked to the direct dimer-singlet phase in the large-$g$ limit, and thus exhibits no such topological entropy term~\cite{grover11}.
Also for quantum critical systems, finite constant contributions have been calculated, which depend e.g. on the boundary conditions~\cite{metlitski09}. However, $d$ may also receive a shape-contribution depending on the aspect-ratio, which then makes a direct interpretation of the actual value of $d$ in terms of possible universal contributions difficult, given that a full characterization of the aspect-ratio dependence has not been achieved yet. Depending on the range of available system sizes, one may also require to include further subleading finite-size corrections to the thus far considered terms, which scale with the inverse length $1/l$.  
\begin{figure}[!t]
\includegraphics[width=\linewidth]{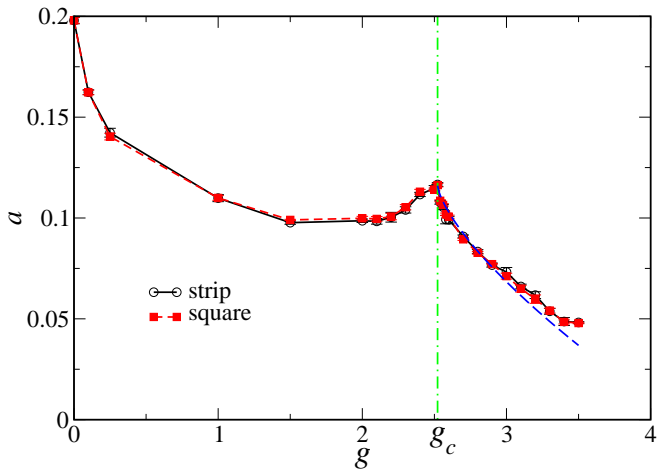}
\caption{(Color online) Area law prefactor $a$ of the second R\'enyi entropy $S_2$ as a function of the coupling ratio $g$ for strip and square shaped subregions. The dashed line is a fit to $a-a(g_c)= r \xi^{D-1}\propto |g-g_c|^\nu$ with the correlation length exponent $\nu=0.71$. }
\label{fig:avsj}
\end{figure}
Due to the finite accuracy of our QMC data, such extended fitting formula in general  result in large uncertainties in the fitted parameters. In previous QMC studies of two-dimensional quantum spin systems on similarly sizes lattices, restricted fitting functions were  employed that suppress such $O(1/l)$ corrections~\cite{kallin11,humeniuk12}. Proceeding along this line, one  has to consider that  the resulting error bars on the fitting parameters, while accounting for statistical uncertainties, do not reflect systematic deviations due to finite size-effects. In the following, we thus compare our QMC results to  calculations performed directly in the thermodynamic limit, e.g. from series expansions, whenever such results are available. 
 \begin{table}
\begin{tabular}{| c || c | c | c | }
   \hline                        
   $g$   & a          & $c^{\mathrm{st}}$  & $c_c$ \\ \hline   
   0      & 0.197(3)   & 1.35(4)  & -0.143(2) \\
   0.25  & 0.142(3)   & 0.78(5)  & -0.051(5) \\
   1      & 0.110(3)   & 0.82(4) & -0.051(3) \\
   2     & 0.980(3)    & 0.70(2) &  -0.057(4) \\
   $g_c$ & 0.117(3)   & $\approx 0$ & -0.062(3) \\
   3     & 0.073(3)   & $\approx 0$ & $\approx 0$ \\
   3.4   & 0.049(3)   & $\approx 0$ & $\approx 0$ \\
   \hline  
\end{tabular}
\caption{Coefficients of the area law and logarithmic contributions to the second R\'enyi entropy  obtained from fitting the QMC data.}
\end{table}

Performing such fits to $S_2= a l + c\ln(l) + d$, we  extract in particular the leading area-law coefficient $a$, which is shown in Fig.~\ref{fig:avsj} as a function of $g$ for both subregion shapes and given in Tab. I  at several values of $g$.  For these fits, the full range of available system sizes $L=8,...,20$ was employed, and the fits returned values of $\chi^2/\mathrm{DOF}=O(1)$. In addition, we performed fits excluding the smaller values of $L$, i.e. for ranges $L=10,...,20$ and $L=12,...,20$. The corresponding values of $a$ were found to agree within error bars with those given in Fig.~\ref{fig:avsj} and Table I, but  they exhibit larger uncertainties.   
Our value for $a=0.197(3)$ at $g=0$ is  in accord within statistical uncertainties with twice the value for the single layer Heisenberg model result reported in previous  studies~\cite{kallin11,humeniuk12}.
We furthermore find from Fig.~\ref{fig:avsj}, that within the statistical uncertainties  the extracted values of $a$ agree among the two different subregion geometries.

Fig.~\ref{fig:avsj} shows that $a$ exhibits a non-monotonous behavior as a function of $g$, and  develops a local maximum at criticality, $g=g_c$. The behavior of $a$ in the vicinity of $g_c$ is consistent with an algebraic scaling of $a-a(g_c)= r \xi^{D-1}\propto |g-g_c|^\nu$ in this two-dimensional systems ($D=2$), where $\nu$ denotes the  correlation length exponent, which in the present case equals that of the three-dimensional Heisenberg model universality class.
A fit of our data for $g\gtrsim g_c$ to the the standard value of the correlation length exponent $\nu=0.710(2)$ (Ref.~\onlinecite{hasenbusch01}) entering the above scaling relation is shown in Fig.~\ref{fig:avsj}, while an unbiased algebraic fit to our data returns a  lower value of $\nu\approx 0.55$.  
However, given the statistical uncertainty on $a$ within the critical region, we do not consider this an reliable independent estimate of the exponent. Nevertheless, we can conclude that $r<0$ on both sides of $g_c$, in accord with general expectations~\cite{casini12}.

We next consider  possible subleading logarithmic contributions to $S_2$. For the strip subregions with smooth boundaries and no corners, a   logarithmic term  is expected from the presence of Goldstone modes in the antiferromagnetically ordered regime: in accord with the numerical observation of subleading logarithmic contributions in the square lattice Heisenberg model~\cite{kallin11},
the authors of Ref.~\onlinecite{metlitski11} put forward a universal subleading logarithmic contribution in systems with spontaneously broken continuous symmetry equal to $c=N_G(D-1)/2$, where $N_G$ denotes the number of Goldstone modes.  In two dimensions, $D=2$, and for the $N_G=2$ SU(2) symmetry breaking within the finite-$g$ region, this leads to a contribution with $c=1$, while at $g=0$, where the symmetry is trivially enhanced to a SU(2) $\times$ SU(2) symmetry due to the decoupling of the layers, we obtain $N_G=4$ and thus twice the above value, i.e. $c=2$. 
\begin{figure}[!t]
\includegraphics[width=\linewidth]{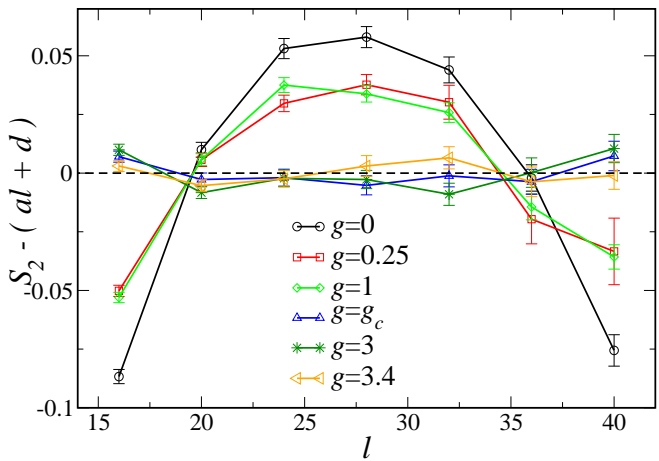}
\caption{(Color online) Residuals to a linear fit $al+d$ of the second R\'enyi entropy $S_2$ of a strip subregion  as a function of the boundary length $l$  at different values of the coupling ratio $g$.}
\label{fig:dev}
\end{figure}

\begin{figure}[!t]
\includegraphics[width=\linewidth]{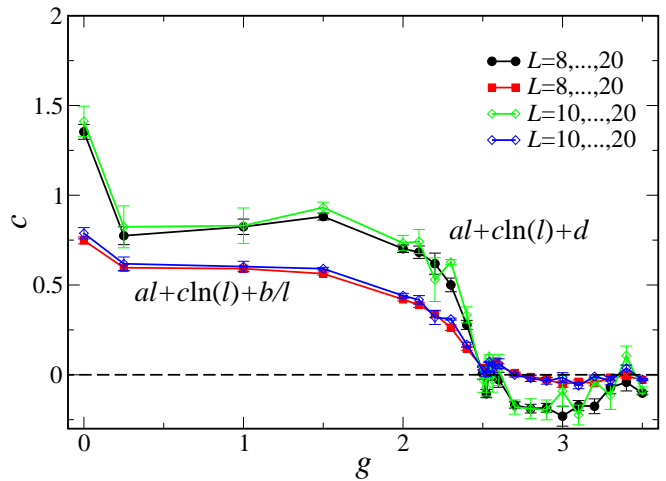}
\caption{(Color online) Coefficient of the logarithmic contribution $c$ to the second R\'enyi entropy obtained from employing the indicated fitting formula to the numerical data for a strip shaped subregion at different values of the coupling ratio $g$.  }
\label{fig:cst}
\end{figure}

 To assess the statistical signature  for possible subleading logarithmic terms with respect to the QMC statistical uncertainties, we  show  in Fig.~\ref{fig:dev} the deviation of $S_2$ for the strip subregion  
to a linear fit $al+d$, i.e.  without a logarithmic term, for several values of the coupling ratio $g$. From this figure, we find that for those values of $g$ with $g<g_c$, such a restricted fit significantly fails to account for the overall $l$-dependence of $S_2$.  For $g>g_c$ and at criticality, $g=g_c$, the QMC data for the strip subregions is consistent with the absence of a logarithmic term, as the residuals fluctuate about zero on a scale of the order of the statistical uncertainties. On the other hand, when fitting the QMC data for the strip subregion to the scaling formula $S_2=al+c \ln(l)+d$, i.e. including a logarithmic term, we obtain the values of $c$ given in Fig.~\ref{fig:cst}.
We show in this figure the results from fitting the data for all system sizes, $L=8,...,20$, as well as a restricted fit for $L=10,...,20$, thus skipping the lowest, $L=8$, data. We obtain for $g<g_c$ finite, positive values of $c$ for both fitting ranges. Close to criticality, the fitted values of $c$ show a spread of values near $c\approx 0$. For $g>g_c$, we obtain small, negative values for $c$; at some values of $g$ within this regime, the fitted values furthermore deviate strongly and even in sign among the two different fitting ranges. This indicates, that those finite, negative values of $c$ result from overfitting to the statistical noise in our data instead of indicating a significant, finite logarithmic term in this regime. Our QMC data thus imply a finite positive value of $c$ for the strip subregions within the $g<g_c$ antiferromagnetically ordered regime, while at $g_c$ and within the dimerized phase, our data is consistent also with the absence of a 
logarithmic contribution, $c=0$.  

Within the antiferromagnetically ordered regime for $g \lesssim 2$, we obtain values of $c$  between 0.7 and 0.8, while for $g=0$ we obtain $c=1.35(2)$, cf. Table I (where the $c$ values for the strip geometry are denoted as $c^\mathrm{st}$). This corresponds to an expected increase of $c$ with respect to the  Goldstone mode counting. However,  both values fall significantly below the expected values of $c=1$ and $c=2$, respectively, which result from the formula of Ref.~\onlinecite{metlitski11}. In previous studies on the single square lattice Heisenberg model, a corresponding value of $c=0.74(2)$ was obtained for the $L/2\times L$ strip geometry~\cite{kallin11}. This value is similar to our values in the finite-$g$ regime  $g \lesssim 2$, in accord with a possibly universal value for the $N_G=2$ SU(2) symmetry breaking. However our value  $c=1.35(2)$ at $g=0$  falls below the value $2\times 0.74(2)=1.48(4)$. Such differences could be the result of finite-size effects, referring to  the difficulty of estimating the logarithmic term based on the still small linear system sizes available in QMC 
studies. 
In fact, compared to Ref.~\onlinecite{kallin11}, where a finite single square lattice was considered up to $L=32$, we effectively simulate at $g=0$ two isolated finite square lattices up to $L=20$ only.   
The  decrease of $c$ seen in Fig.~\ref{fig:cst} upon increasing $g$ and approaching the quantum critical point suggests another manifestation of  finite-size effects, since the asymptotic behavior will be accessible only for subregions that extend well beyond the correlation length scale. It is  difficult to overcome these problems based on finite-system approaches, given that the correlation length scale becomes exceedingly large in this regime. 

Figure~\ref{fig:cst} also shows the value of $c$ that results when fitting the QMC data to the function $S_2=a l +c \ln(l) + b/l$. Such a functional form, with $c=0$, is expected to apply within the quantum disordered region, given (i) the absence of a topological entanglement entropy in this trivial gapped phase ($d=0$) and (ii)  the inclusion of the leading finite-size correction term ($b/l$), consistent with general arguments~\cite{grover11}. Our data in the regime $g>g_c$ fits well to this formula (with values of $\chi^2/\mathrm{DOF}=O(1)$), returning very small values of $c$ that are, given the statistical uncertainty, consistent with $c=0$. Inside the antiferromagnetically ordered phase, this fitting formula returns finite values of $c$ that fall  below the values obtained from the above fit that  allowed for a finite $d$ term. However, inside the ordered region,  a finite value of $d$ is expected in general due to the aspect-ratio-dependence observed in previous work on the square lattice~\cite{
kallin11} and its inclusion is thus required in this region. We  thus also  attempted to fit the strip subregion data to a function containing both a constant $d$ as well as the $b/L$ finite-size correction term. However,  such a combined  fitting ansatz results in large uncertainties on the fit parameters, since within the finite range of the accessible lattice sizes, the numerical data is not constraining sufficiently all these fit parameters, given to the finite QMC statistical uncertainties. 

Finally, we consider  logarithmic contributions to the entanglement entropy scaling that relate to the $90^{\circ}$ corners in the subregion boundary of the square subregions. One means of extracting such corner contributions would be to perform separate fits of the numerical data to the scaling formula  $S_2=al+c \ln(l)+d$ for both subregion shapes, and to then compare the obtained values for $c$ between the two cases. However, we find that such an approach is strongly affected by the combined uncertainties in fitting the values of $c$ for the two subregion shapes.  Instead, we consider here directly the difference $\Delta S_2$ between the two entropies, cf. Eq.~(\ref{eq:ds2}), which is shown for various  values of $g$ in Fig.~\ref{fig:DSvsl}. For the values of $g>g_c$, the data for $\Delta S_2$ shows no significant $l$-dependence with respect to the statistical uncertainties.  For the other considered cases,  the behavior of $\Delta S_2$ is consistent within the statistical uncertainties with a linear dependence on $\ln(l)$, which results if the 
separate entropies for both subregion shapes independently follow an $S_2=al+c \ln(l)+d$ scaling with the same value of $a$ for both strip and square subregions, as supported by our data in Fig.~\ref{fig:avsj}. Assuming that other contributions to the logarithmic term, i.e. those from the Goldstone modes are equal for both subregions (this assumption is in accord with the formula in Ref.~\onlinecite{metlitski11}, and is not required if such contributions are absent, e.g. at the quantum critical point), we may thus extract the residual corner contribution $c_c$  by a linear regression of the data shown in Fig.~\ref{fig:DSvsl}, i.e. by fitting to $\Delta S_2= c_c \ln(l)+\Delta d$. The resulting values of $c_c$ are collected in Table I. 
We find that at $g=0$, the resulting value of $c_c=-0.143(2)$ is lower than twice the value $-0.10(2)$ obtained using a valence bond QMC approach in Ref.~\onlinecite{kallin11}, while within error bars is it consistent with the Ising series expansion result $-0.080(8)$ taken from the same reference. The values of $c_c$ that we extract for the finite-$g$ antiferromagnetically ordered  region are very similar.  Finally, we note that the value $c_c=0.062(3)$ that we obtain at $g=g_c$ implies a contribution of $-0.016(1)$ for each single $90^{\circ}$ corner at the quantum critical point. This value agrees well with the recent result obtained in Ref.~\onlinecite{kallin14} from a numerical linked cluster calculation, and deviates -- even when  considering the QMC statistical uncertainty -- with the corresponding value for a three-component Gaussian theory, $3\times (-0.0064)=-0.0192$, based  on the results of Ref.~\onlinecite{casini07}. The difference between these values however is not very large. This is in 
accord with the fact that while the true quantum critical behavior is controlled by the Fisher-Wilson fixed point, the two theories are connected by an $\epsilon$-expansion about four dimensions.

\section{Discussion}
We employed quantum Monte Carlo simulations to analyse the scaling of the entanglement entropy in the square lattice bilayer Heisenberg model. For the dominant area-law term, we exhibited the enhancement of this non-universal contribution at the quantum critical point. Furthermore, we find that it scales consistently with general scaling considerations in the vicinity of the quantum critical point.

In addition, we examined subleading terms  that contribute a logarithmic scaling with the boundary length. 
Our numerical data is consistent with the absence of a logarithmic contribution to the entanglement entropy scaling for a subregion with a smooth boundary both at the quantum critical point and in the quantum disordered region. In the antiferromagnetically ordered region, we extracted a finite logarithmic term which can be accounted for by the presence of Goldstone modes in this region. In particular, in the limit of decoupled layers, the observed increase of the logarithmic coefficient is in accord with the enhanced  symmetry in this limit, which leads to a doubling of the number of Goldstone modes. 
As in previous numerical studies, the estimated scaling coefficients however fall below the expected values from the Goldstone mode counting formula in Ref.~\onlinecite{metlitski11}
We cannot exclude that this deviation  is  due to finite-size corrections in the scaling coefficients, which may also be indicated by the suppression of the  coefficient upon approaching the quantum critical point. 
To settle this issue, it  will be important to calculate the Goldstone-mode contribution by methods less susceptible to finite-size effects. 

We also estimated the subleading logarithmic contribution that is expected due to  the presence of $90^{\circ}$ corners in the subregion boundary. At  criticality, this contribution was conjectured  to be universal in that it provides a measure of the number of components of the effective field-theory describing the quantum critical point~\cite{kallin14}.  Our estimated value of this corner term from quantum Monte Carlo simulations is consistent with the recent result from a numerical linked-cluster expansion, which was performed directly in the thermodynamic limit and at zero temperature~\cite{kallin14}.
As the authors of Ref.~\onlinecite{kallin14} point out, this value is essentially three times the value that is obtained for the corner term at the quantum critical point of the two-dimensional  transverse field Ising model. This indeed supports the idea that this corner term provides a universal count of the degrees of freedom at the quantum critical point, analogous to the central charge in two-dimensional conformal field theory. To strengthen this conclusion, it will be important to assess the universality of this result by analyzing other  two-dimensional quantum critical points of well known universality. One possible route will be the study of differently dimerized quantum spin systems or models with an $XY$ symmetry. It may also be interesting to assess, if the value of the corner term is invariant with respect to e.g. enhanced irrelevant scaling variables, such as those proposed to emerge from three-triplon processes in the staggered-dimerized spin-model~\cite{fritz11}. 

Inside the gapped, quantum disordered regime, our numerical data is consistent with the absence of a constant term in the entanglement entropy, in accord with the absence of topological order in this model.
We however did not analyse in further detail the constant contribution to the entanglement entropy beyond the quantum disordered regime.  This would require to separate an inherent constant term (e.g. at the quantum critical point) from  additional constant terms that were seen to arise in studies on torodial simulation cells due to the aspect-ratio dependence~\cite{kallin11, inglis13}.  Further work will be required to entangle these contributions, e.g. by establishing the functional form of the aspect-ratio dependence.

\acknowledgements

We have profited from discussions with T. Grover, R. G. Melko, T. Roscilde and S. Trebst. Financial support by the DFG under Grant WE 3649/3-1 is gratefully acknowledged, as well as the allocation of CPU time within JARA-HPC, the CHEOPS cluster at the  Universit\"at zu K\"oln, and  from JSC J\"ulich.

%
%
\end{document}